\documentclass[12pt]{article}

\usepackage{graphicx}
\usepackage{subfigure}
\usepackage{amsmath}
\usepackage{ulem}
\usepackage[usenames,dvipsnames]{color}
\usepackage[bookmarks=true,bookmarksopenlevel=0,bookmarksnumbered=true,hypertexnames=false,
   colorlinks,linkcolor={blue},citecolor={blue},urlcolor={red},
   pdfstartview={FitV},unicode,breaklinks=true]{hyperref}
\usepackage[small]{caption}

\definecolor{orange}{RGB}{160,82,45}

\newcommand{\etal}{\textit{et al}}
\newcommand{\pop}[1]{\textit{Phys. Plasmas.} \textbf{#1}}
\newcommand{\rsi}[1]{\textit{Rev. Sci. Instrum.} \textbf{#1}}

\newcommand{\prl}[1]{\textit{Phys. Rev. Lett.} \textbf{#1}}
\newcommand{\ppcf}[1]{\textit{Plasma Phys. and Contr. Fusion} \textbf{#1}}
\newcommand{\pr}[1]{\textit{Phys. Rev.} \textbf{#1}}
\newcommand{\ft}[1]{\textit{Fusion Technology} \textbf{#1}}
\newcommand{\fst}[1]{\textit{Fusion Science and Technol.} \textbf{#1}}
\newcommand{\fed}[1]{\textit{Fusion Eng. and Design} \textbf{#1}}
\newcommand{\nf}[1]{\textit{Nucl. Fusion} \textbf{#1}}

\begin{document}

\vspace{5mm}
\begin{center}
\LARGE{\bf
Electron Bernstein waves emission in the TJ--II Stellarator
\\[5mm]}
\large{
J. M. Garc\'i{a}-Rega\~{n}a$^1$, \'{A}. Cappa$^1$, F. Castej\'{o}n$^1$,\\ J.B.O Caughman$^2$,
M. Tereshchenko$^{3,4}$, A. Ros$^1$, \\ D. A. Rasmussen$^2$ and J. B. Wilgen$^2$
\\[3mm]}
\small{
${}^1$ Laboratorio Nacional de Fusi\'{o}n, CIEMAT, 28040, Madrid, Spain\\[2mm]
${}^2$ Oak Ridge National Laboratory, Oak Ridge, TN USA\\[2mm]
${}^3$ Prokhorov General Physics Institute,\\[-0em] Russian Academy of Sciences, Moscow, Russia\\[2mm]
${}^4$ BIFI Instituto de Biocomputaci\'on y F\'isica\\[-0em] de Sistemas Complejos, Zaragoza, Spain
}\\[2mm]
\small{email: josemanuel.garcia@ciemat.es}
\end{center}

\begin{abstract}
Taking advantage of the electron Bernstein waves heating (EBWH) system of the TJ--II 
stellarator, an electron Bernstein 
emission (EBE) diagnostic was installed. Its purpose is 
to investigate the B--X--O radiation properties in the zone where optimum 
theoretical EBW coupling is predicted.
An internal movable mirror shared by both systems allows to 
collect the EBE radiation along the same line of sight that is used for EBW heating.
The theoretical EBE has been calculated for different orientations of the internal 
mirror using the TRUBA code as ray tracer. A comparison with 
experimental data obtained in NBI discharges is carried out. The 
results provide a valuable information regarding 
the experimental O--X mode conversion window expected in the 
EBW heating experiments. Furthermore, the characterization of the 
radiation polarization shows evidence of the underlying 
B--X--O conversion process.
\end{abstract}

% SECTION BLOCK SECTION BLOCK SECTION BLOCK SECTION BLOCK SECTION BLOCK SECTION BLOCK
% SECTION BLOCK SECTION BLOCK SECTION BLOCK SECTION BLOCK SECTION BLOCK SECTION BLOCK
% SECTION BLOCK SECTION BLOCK SECTION BLOCK SECTION BLOCK SECTION BLOCK SECTION BLOCK
\section{Introduction}
Plasma heating by Electron Bernstein waves \cite{Bernstein_pr_109.1_1958}
has been successfully demonstrated in several 
magnetic confinement devices \cite{Mueck_prl_98_2007,Laqua_ppcf_41_1999,Igami_nf_49_2009,Preinhaelter_ppcf_51_2009}. 
The main advantage of this technique is that it offers the possibility to heat 
plasmas above the density limits of the standard  electromagnetic modes.
For a complete review of the EBW topic see \cite{Laqua_ppcf_41_2007} 
and references therein.
Moreover, the emitted Bernstein waves from the over-dense plasma core
can be helpful for the plasma diagnosis in high density scenarios.
See, for instance, ref. \cite{Volpe_rsi_74.3_2003}, where EBE is used as an electron temperature 
diagnostic and ref. \cite{Jones_pop_11.3_2004} where it is employed to measure the magnetic field.

The TJ--II stellarator \cite{Alejaldre_ft_17_1990} is a medium size heliac 
with major radius $R_0=1.5$ m, minor radius $a\approx 0.2$ m and magnetic field strength 
on axis around $0.95$ T. In this device, the use of the available ECR heating 
sources --two gyrotrons delivering 300 kW power each 
and injecting X--mode polarized waves at second harmonic ($f=53.2$ GHz)-- 
is limited by the cut-off density for that frequency and that
mode, i.e. $n_e\approx 1.7\times 10^{19}$ m$^{-3}$. 
Two neutral beam injectors allow to heat the plasma above 
this limit and therefore, additional ECRH heating in NBI plasmas 
becomes possible only through electron Bernstein waves.
The heating scheme using Bernstein waves for TJ-II is based on the O--X--B double mode conversion using low field side
launching \cite{Castejon_fst_52.2_2006}.

As it is well-known, the efficiency of the O--X mode conversion 
depends strongly on the launching direction and the polarization 
of the incident wave. Thus, in order to find an optimum experimental 
heating efficiency, a steerable mirror was installed inside 
the vacuum vessel, allowing a horizontal and vertical scan of 
the ECRH beam launching direction around the direction for which 
maximum O--X mode conversion efficiency is expected. Note that a 
perfect optimization procedure should also 
include the displacement of the mirror center, since for each 
penetration location of the plasma wave a different well 
defined injection direction is required. This is not possible with the existing design.       

In the present setup, the mirror can also be positioned so that the plasma 
radiation coming along the EBWH launching direction is redirected towards a 
radiometer antenna that measures thermal emission at 28 GHz. 
With this setup, measuring and heating can not be
performed at the same time.
Since both the EBWH and the ECE/EBE systems share the same line of sight, 
the analysis of electron cyclotron radiation in plasmas 
above the O mode cut-off density for 28 GHz ($n_e\approx 1.0\times 10^{19}$ 
m$^{−3}$) may provide a very valuable information 
to determine the optimum direction for O--X--B heating. As a matter of fact, 
because of the low cut-off densities of the 28 GHz 
electromagnetic modes, 
any radiation detected in plasmas above this density must come from
an inverse B--X--O mode conversion unless the contribution of the peripheral underdense
plasma to the emission is high enough to mask the B--X--O component. This point will
be addressed in the discussion.

In this work, we present the theoretical characterization of the 
EBE resulting from B--X--O conversion,
for different orientations of 
the launching mirror and different rotation angles 
of the radiometer, in the vicinity of the position that provides 
the maximum theoretical O--X conversion efficiency. The numerical results are 
obtained with the ray tracing code TRUBA and the well-known 
concepts in traditional ECE calculations, i.e. energy balance between emission 
and absorption as well as black body approximation. 
The theoretical results are compared with the experimental ones, obtained 
under ECRH + NBI discharges. 

The article is outlined as follows: section \ref{systems} focuses on the description of the
experimental setup; section \ref{num_results} deals with the numerical EBE calculations; 
section \ref{exp_results} is devoted 
to the experimental results and their comparison with the predictions 
presented in section \ref{num_results}; and finally, section \ref{conclusions} 
summarizes and discusses the main results. 
The appendix describes the calculation of the expected power 
in the different polarizations measured by the radiometer when it is rotated around its symmetry axis.

% SECTION BLOCK SECTION BLOCK SECTION BLOCK SECTION BLOCK SECTION BLOCK SECTION BLOCK
% SECTION BLOCK SECTION BLOCK SECTION BLOCK SECTION BLOCK SECTION BLOCK SECTION BLOCK
% SECTION BLOCK SECTION BLOCK SECTION BLOCK SECTION BLOCK SECTION BLOCK SECTION BLOCK
\section{EBWH launching system and EBE diagnostic}
\label{systems}
The EBWH launching system consists of a corrugated
waveguide that transmits up to 300 kW of microwave power into the vacuum
vessel, where a steerable ellipsoidal mirror ($170$ mm $\times 190$ mm) 
focuses the beam and defines the launching direction. Prior to coupling the 
radiation to the waveguide, two elliptical focusing mirrors and two $\lambda/4$ 
and $\lambda/8$ plane polarizers provide a Gaussian beam with the properly
\begin{figure}[h]
\begin{center}
\includegraphics[width=0.75\textwidth]{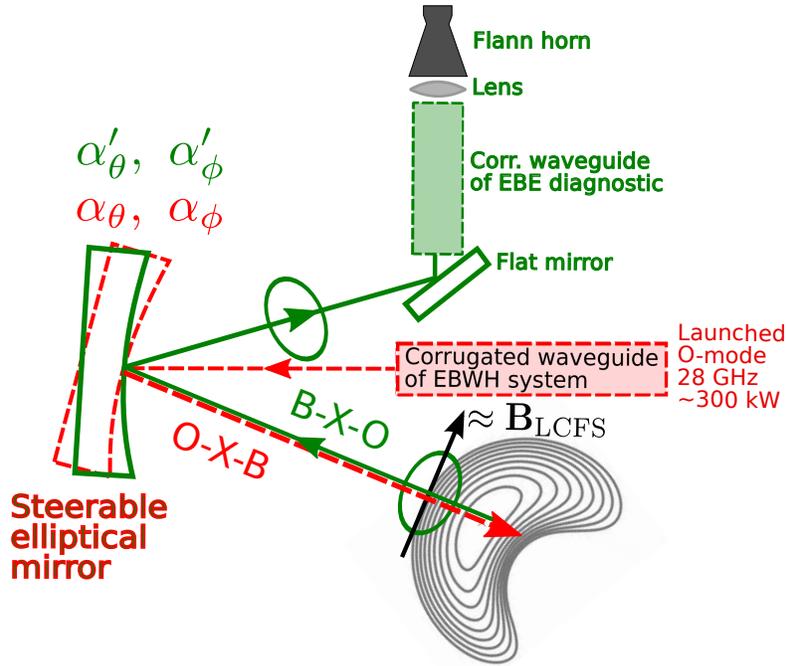}
\caption{EBWH launching system and EBE diagnostic. Different mirror 
positioning angles are used for EBW heating ($\alpha_{\phi}$, $\alpha_{\theta}$) and 
for EBE detection ($\alpha_{\phi}^\prime$, $\alpha_{\theta}^\prime$). 
Radiated power suffers two consecutive reflections and is coupled to 
the detection waveguide.}
\label{fig1:systems}
\end{center}
\end{figure}
polarized electromagnetic field. For the details of the EBW heating system see 
refs. \cite{Curto_fed_74_2005,Castejon_nf_48.7_2008}. The position of the movable mirror 
is determined by two orientation angles along the horizontal (or toroidal) and vertical (or poloidal) 
directions, $\alpha_{\phi}$ and $\alpha_{\theta}$ respectively. If the launching 
direction is given by these angles, the %28 GHz 
radiation coming from this same 
direction can be redirected to the 28 GHz heterodyne radiometer 
using a different pair of angles $\alpha_{\phi}^\prime$ and $\alpha_{\theta}^{\prime}$ 
and a flat mirror attached to a corrugated waveguide, 
similar to the one used in the launching system (see figure \ref{fig1:systems}). 
After its passing through a glass 
focusing lens located outside the vacuum chamber, the radiation is detected 
by a quad-ridged dual-polarized microwave horn. The lens modifies the beam pattern 
at the output of the waveguide in order to obtain an efficient coupling to the horn antenna. 
For further details of the EBE diagnostic see \cite{Caughman_ec14_2006, Caughman_fst_57_2010}.

Once the 28 GHz radiation reaches the horn, it is split 
into two separate components according to two
 orthogonal detection directions. Properly calibrated, the sum 
of both signals, $I_{\text{\tiny{EBE1}}}$ and $I_{\text{\tiny{EBE2}}}$, 
corresponds to the total radiative temperature $T_{\text{\tiny{rad}}}$.
Accordingly to the properties of an obliquely propagating O mode, 
an elliptical polarization is expected. Therefore, in order 
to perform an experimental measurement of this polarization, 
the orientation of the detection directions may be changed from 
shot to shot by rotating the diagnostic an angle $\zeta$ around 
its symmetry axis. Actually, only the amplitude of the field components 
can be measured to some degree of accuracy and no information about 
the interacting phase between components and thus about the rotation sense 
of the polarization can be extracted. 

Using purely geometrical calculations with the vacuum field configuration 
it can be seen that for $\zeta=10^{\circ}$, channel 2 (EBE2) is aligned 
along the theoretical major axis of the expected O mode polarization ellipse.
Channel 1 is aligned along the minor axis of 
the ellipse.
See the Appendix for a more detailed description of 
the diagnostic rotation and the expected polarization.

% SECTION BLOCK SECTION BLOCK SECTION BLOCK SECTION BLOCK SECTION BLOCK SECTION BLOCK
% SECTION BLOCK SECTION BLOCK SECTION BLOCK SECTION BLOCK SECTION BLOCK SECTION BLOCK
% SECTION BLOCK SECTION BLOCK SECTION BLOCK SECTION BLOCK SECTION BLOCK SECTION BLOCK
\section{EBE numerical simulations}
\label{num_results}

\begin{figure}[t]
\begin{center}
\includegraphics[width=0.60\textwidth]{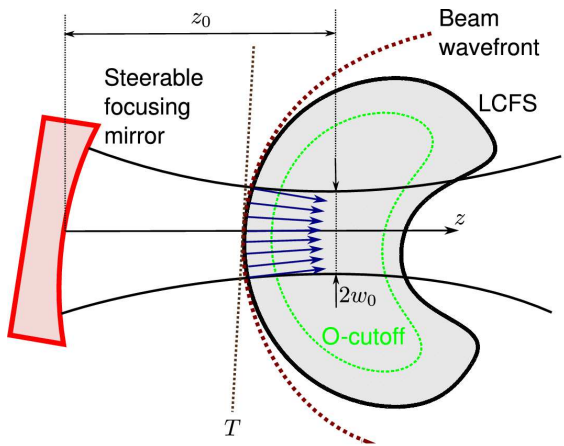}
\caption{(a) The rays are launched perpendicularly to the wave front surface at the plasma edge.
} 
\label{fig:fig3}
\end{center}
\end{figure}

The calculation of the B--X--O radiation has been performed with ray tracing 
simulations using the TRUBA code, whose detailed
description can be found in refs \cite{Tereshchenko_eps30_2003}
and \cite{Castejon_nf_48.7_2008}.

From now on, the internal mirror angles that we set to
collect the 28 GHz radiation will be referred as $\alpha_{\phi}$ and $\alpha_{\theta}$, 
dropping the primes.

To simulate the emission, we need first the ray trajectories obtained 
by tracing a beam launched from the corrugated detection antenna towards 
the plasma (all the simulations have been done launching 121 rays, and for
the standard equilibrium configuration, provided by VMEC-based libraries \cite{Tribaldos_lib}.
The maximum averaged $\beta$ never exceeds 1\%, thus no significant 
modification in the equilibrium is produced, since TJ-II is designed 
to present a very small Shafranov shift.
The ellipsoidal mirror has been designed to focus the beam generated by 
the launching antenna. The incidence angles and the distance to the 
detection waveguide for EBE differ from the ones used in the EBW heating 
configuration and therefore, the mirror is no longer optimized for the 
EBE measurements. The changes that this lack of optimization may 
produce are neglected here. Moreover, note that the ray tracing method
has a fundamental 
limitation when it is applied to the O--X conversion process since it 
cannot take into account the beam spectrum in a fully self-consistent way. 
Therefore, the total O--X conversion efficiency obtained from adding the 
different contribution of each ray --properly weighted with 
the Gaussian beam profile-- differs in general from the O--X conversion efficiency of the 
real electromagnetic field of the beam \cite{Kohn_ppcf_50.1_2008}.
We will discuss this point together with the interpretation of 
the emission data when we come to the summary.

\begin{figure}[t]
\begin{center}
\includegraphics[scale=0.3]{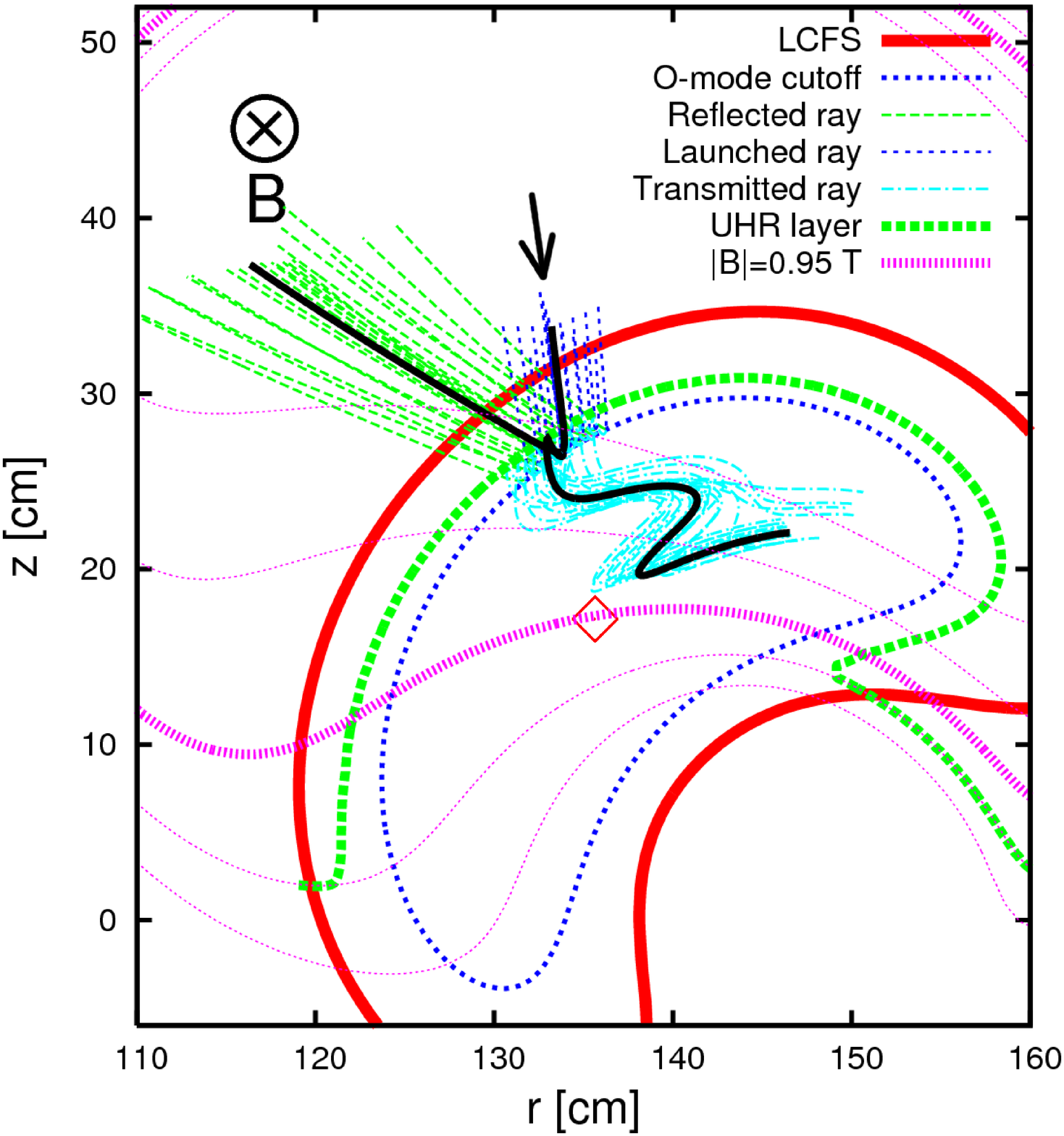}
\caption{Ray tracing with 121 rays (only 21 rays are represented). 
The beam transmission 
efficiency is around 90\%. The black solid lines correspond to the central 
trajectories of the launched, transmitted and reflected rays (see legend).}
\label{fig:fig4}
\end{center}
\end{figure}
Figure \ref{fig:fig4} shows the result of a typical ray tracing simulation. The electron 
density and a temperature profiles used are
fitted to the Thomson Scattering (TS) ones obtained in NBI-heated TJ-II plasmas.  
To calculate the electron Bernstein emission, the radiative transfer equation \cite{Bekefi_1966} 
is solved along the trajectory of the O--X transmitted rays. 
The solution to this equation provides the emission 
intensity per unit frequency, which is given by the well-known expression

\begin{equation}
I_{\omega}=\frac{\omega^2}{8\pi^3c^2}\int_0^{\tau(B)}T_e(\tau)e^{-\tau}\text{d}\tau
\label{eq:eq1}
\end{equation}
where $T_{e}$ is expressed in energy units, $c$ is the speed of light, 
$\omega=2\pi f$ is the wave angular frequency and $\text{d}\tau=-\alpha_{\omega} d\sigma$ is 
the optical thickness, with $\alpha_{\omega}$ the absorption coefficient 
and $\sigma$ the ray coordinate. For each transmitted ray, 
the integral in eq. \ref{eq:eq1} is performed along the ray path up 
to the inner boundary of the conversion layer, i.e. the SX--mode cut--off,
in order to obtain the emission intensity 
$I_i$ prior to the single ray X--O tunneling process. Taking 
into account the conversion efficiency for each particular ray, 
$\eta_i$, and the Gaussian beam profile weighting factor $g_i$, we may 
write the total emission intensity as

\begin{equation}
I_{\text{\tiny{EBE}}}=\sum_i^{n_c}\eta_i g_iI_i 
\end{equation}
where $i$ is the ray index and $n_c$ is the total number of converted rays. 

\subsection{Dependence on mirror position, density and temperature}

\begin{figure}[t]
\begin{center}
\subfigure[]{\label{fig5a}\includegraphics[width=0.48\textwidth]{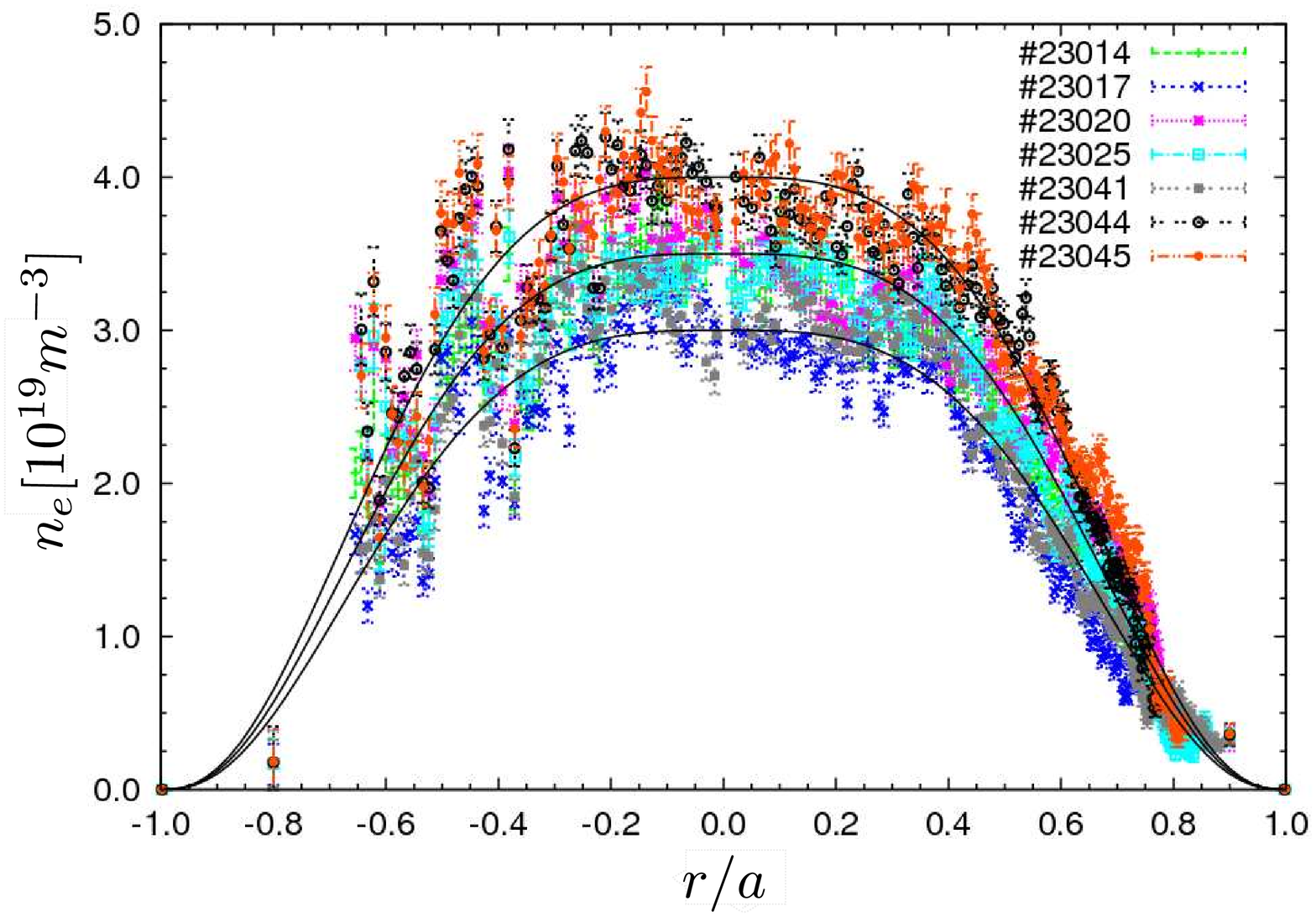}}
\subfigure[]{\label{fig5b}\includegraphics[width=0.48\textwidth]{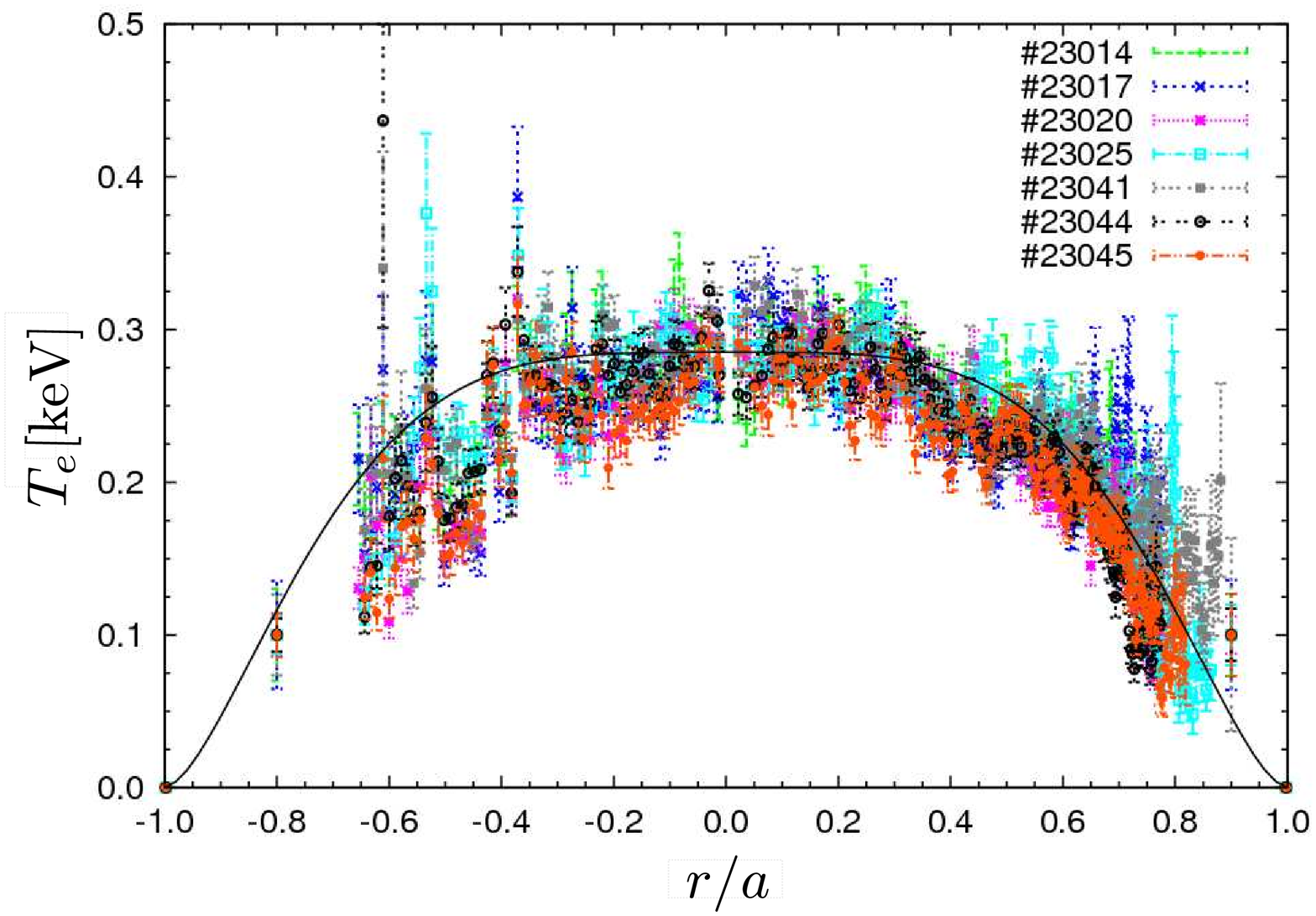}}
\caption{Experimental electron density (a) and temperature (b) Thomson scattering 
profiles obtained in NBI heated plasmas. The numerical calculations have been carried out 
using a fit to the average shape of these profiles.}\label{fig:fig5}
\end{center}
\end{figure}

\begin{figure}[t]
\begin{center}
\subfigure{\includegraphics[width=0.35\textwidth,angle=-90]{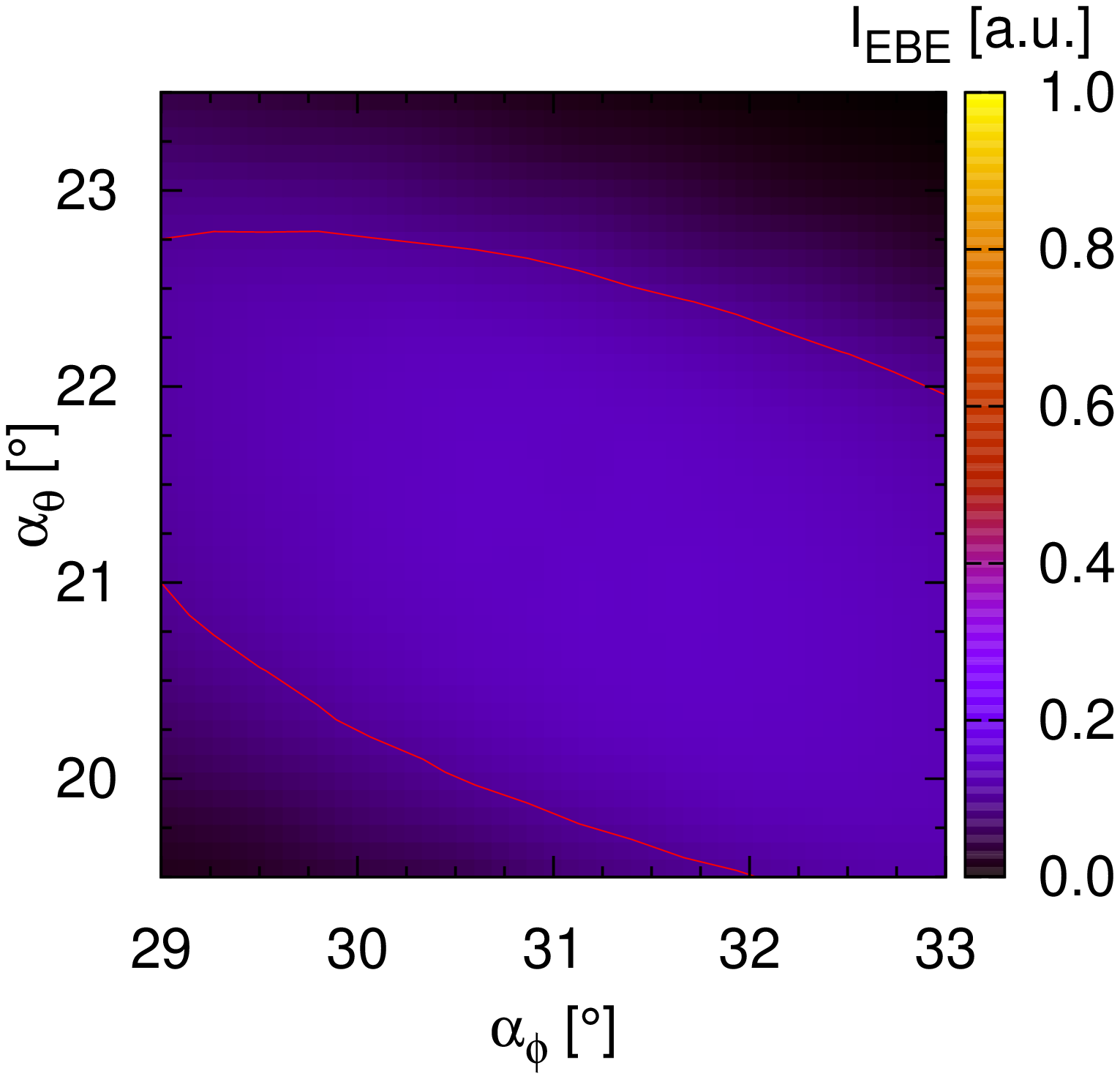}}~(a)
\subfigure{\includegraphics[width=0.35\textwidth,angle=-90]{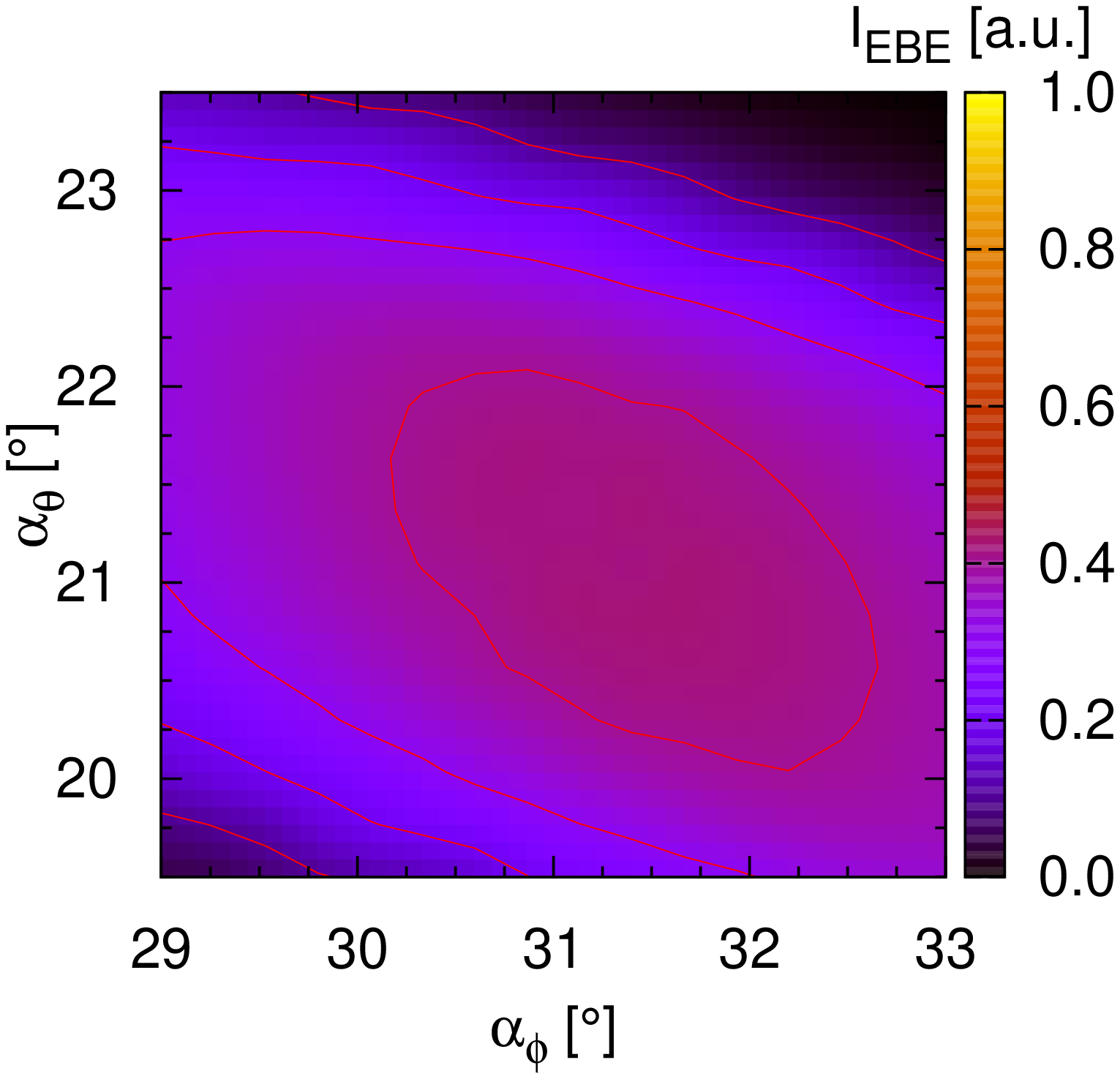}}~(b)
\subfigure{\includegraphics[width=0.35\textwidth,angle=-90]{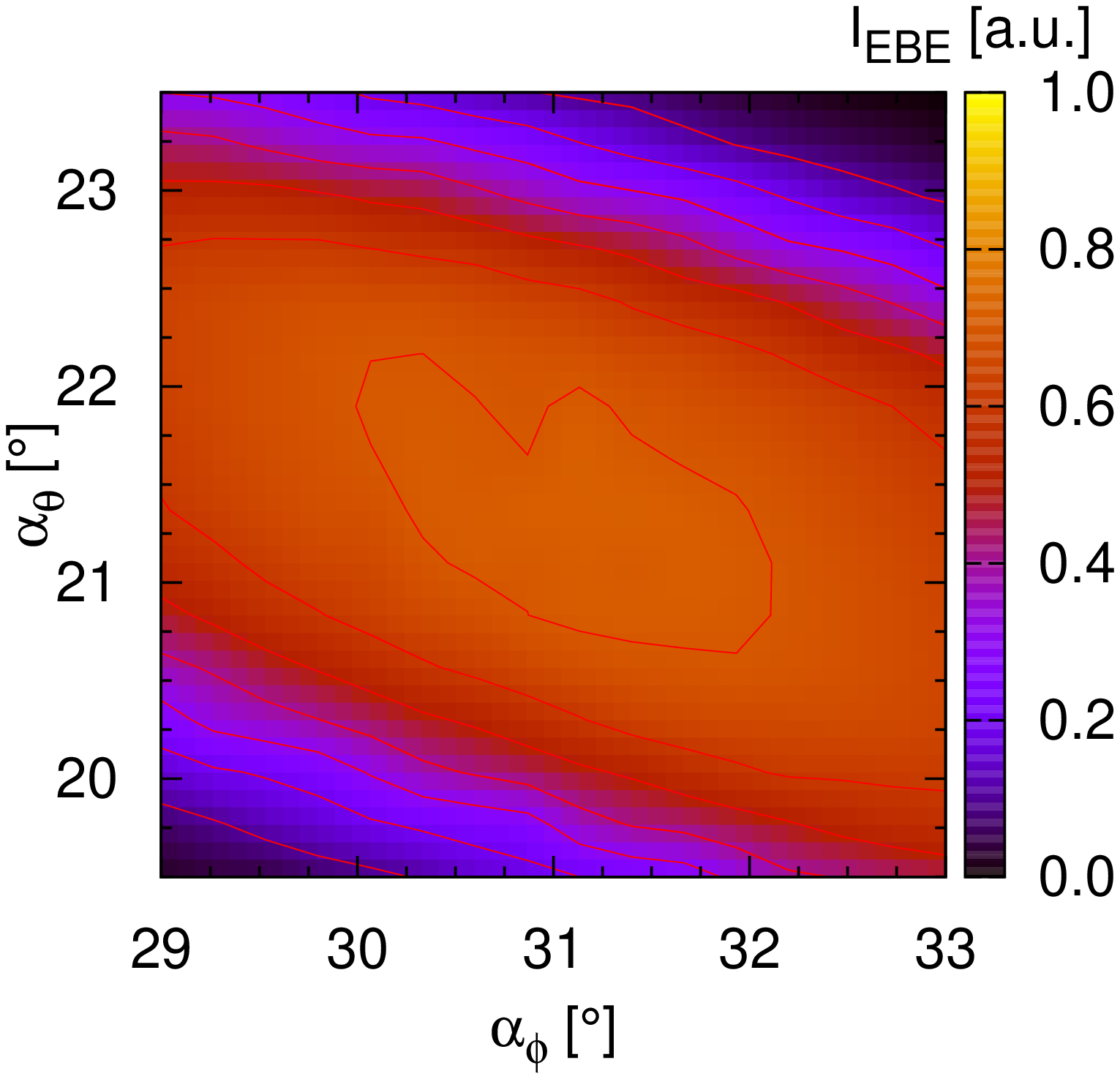}}~(c)
\subfigure{\includegraphics[width=0.35\textwidth,angle=-90]{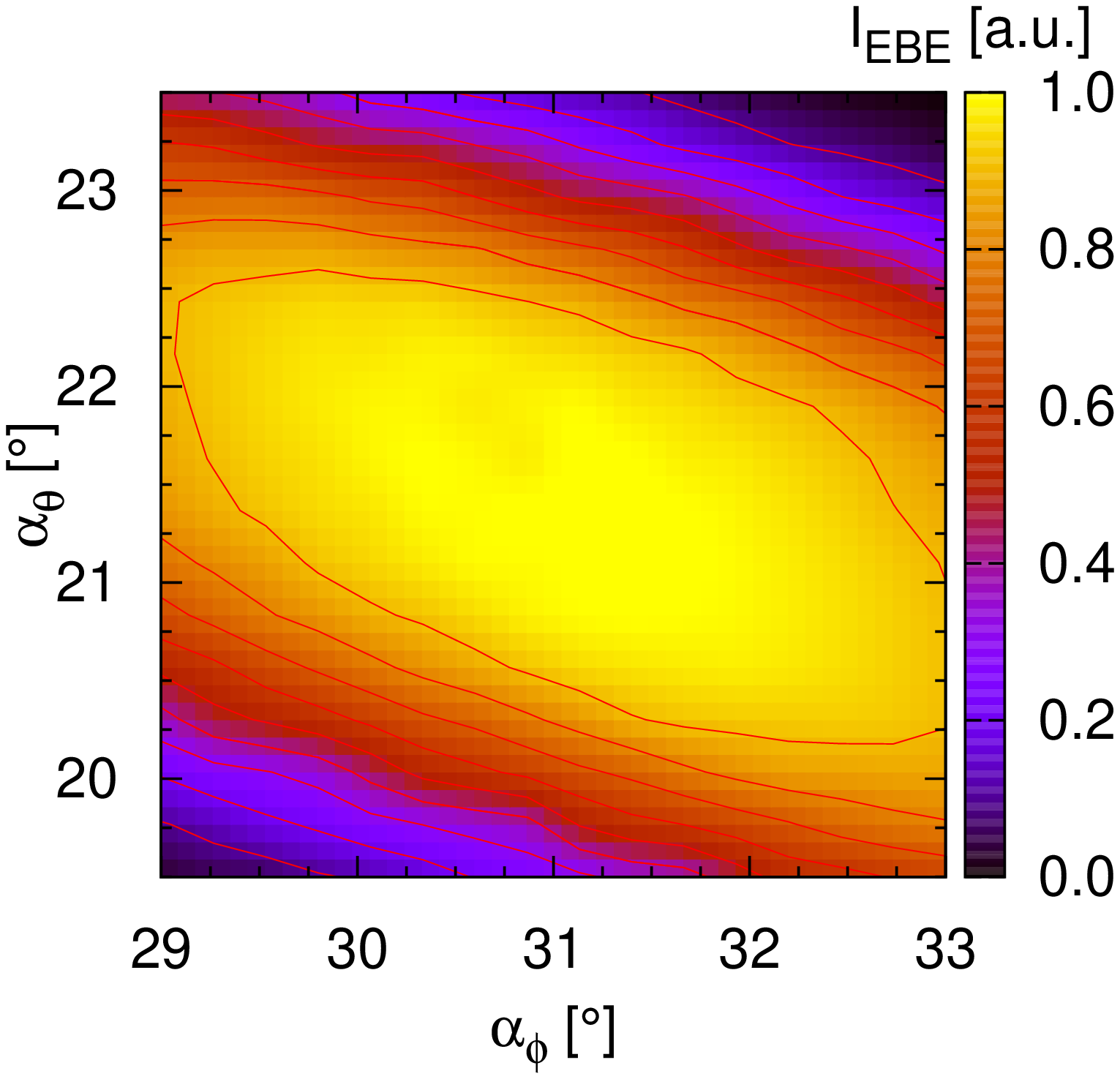}}~(d)
\caption{Dependence of the 28 GHz B--X--O mode converted radiation intensity for 
(a) $T_0=0.1$, (b) $T_0=0.3$, (c) $T_0=0.5$  and (d) $T_0=0.7$ KeV. In all the cases, a central density given by 
$n_0=3.5\times 10^{19}$ m$^{-3}$ has been used. Each map represents the result obtained with 256 launched beams 
and 121 rays per beam.}\label{fig:fig6}
\end{center}
\end{figure}

In order to compare with the result of the experiments presented in section \ref{exp_results}, 
the EBE calculations have been carried out for different lines of sight.
The experimental TS
profiles have been fitted to the functions $n_e(\rho)=n_0(1-\rho^{2n_1})^{n_2}$ 
and $T_e(\rho)=T_0(1-\rho^{2t_1})^{t_2}$, where the set of parameters  
{$\{n_1,n_2,t_1,t_2\}=\{1.6,2.7,2.0,1.7\}$} is found to match best the whole 
set of experimental profiles. A wide range of 
the $n_0$ and $T_0$ values have been considered.  
Figure \ref{fig:fig5} shows some of the experimental  
and analytical electron density (a) and temperature (b) profiles.
The dependence of the electron Bernstein emission on the mirror positioning
angles is plotted in the figures \ref{fig:fig6} for a fixed
central density and four different values of central temperature $T_{0}$.
As expected, the emission intensity increases with 
the electron temperature. In all cases, the maximum emission is located around $\alpha_{\phi}\approx 31^{\circ}$ 
and $\alpha_{\theta}\approx 21.5^{\circ}$.

It is observed that the emission finds a 
stronger dependence on $\alpha_{\theta}$ than on $\alpha_{\phi}$. This is due
to the more pronounced curvature of the plasma along the poloidal direction. 
On the one hand, this makes the incidence direction at the conversion layer deviate 
more sensitively from its optimum $\alpha_{\theta}$ value than from the $\alpha_{\phi}$ one, 
which affects to the conversion efficiency. And on the other hand, it changes appreciably the 
optical path of the rays and consequently the radiation reached at the O cut--off
layer previous to conversion.

\begin{figure}[h]
\begin{center}
\includegraphics[width=0.65\textwidth]{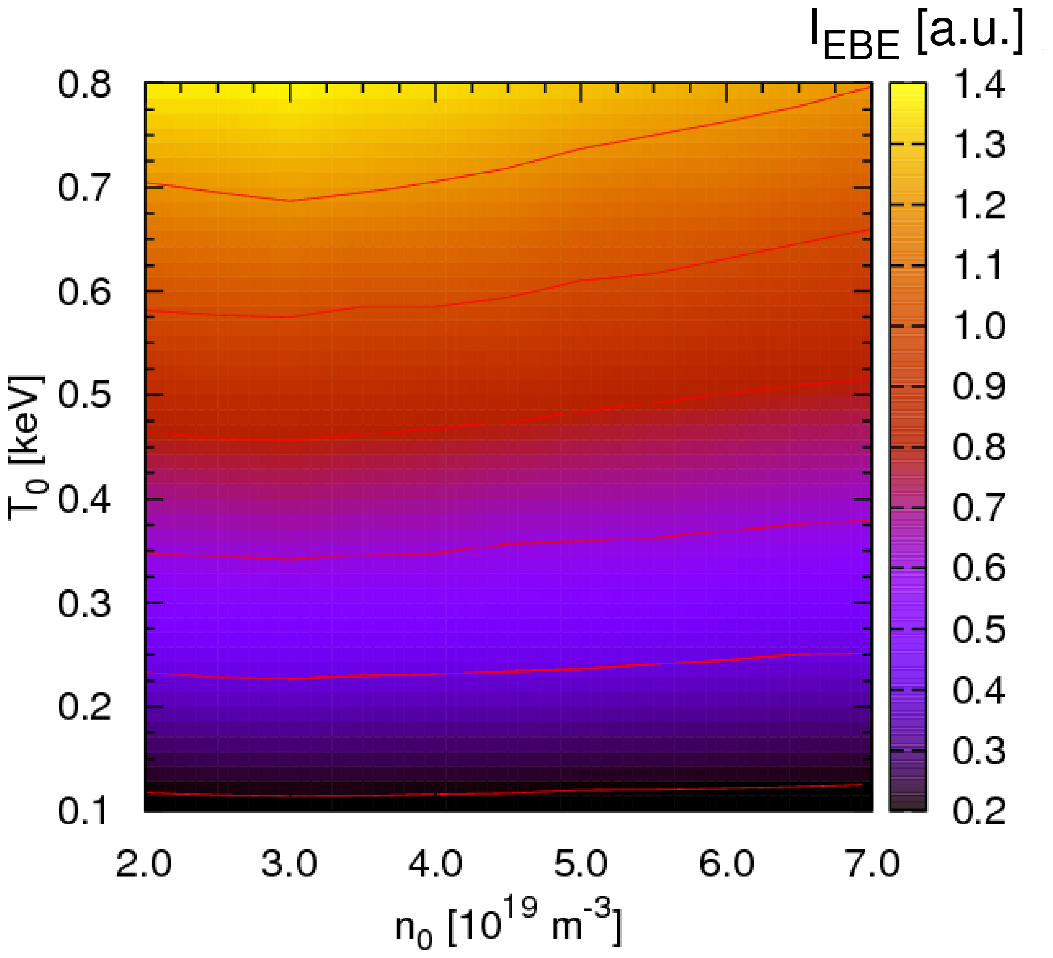}
\caption{Dependence of the B--X--0 radiation intensity on $n_0$ and $T_0$ 
for the line $(\alpha_{\phi},\alpha_{\theta})=(31^{\circ}, 21.5^{\circ})$.}
\label{fig:fig9}
\end{center}
\end{figure}

Regarding the dependence of EBE on central density, note that, as $n_0$ increases, the O--mode cut-off layer moves radially outwards. This modifies the O--X conversion efficiency \cite{Mjolhus_jpp_31_1984} through its dependence on the 
values at the O mode cut-off of the characteristic density gradient scale length ($L_{n}=n_{e}/|\nabla n_{e}|$), $N_{\|}$ and $N_{\bot}$, and the magnetic field strength $B$. 
The final combination of these effects is illustrated in figure \ref{fig:fig9},
where, independently of the central temperature $T_0$, a maximum intensity 
is observed always around $n_0=3\times10^{19}$ m$^{-3}$.

\subsection{Dependence on the rotation angle $\zeta$}

\begin{figure}[t]
\begin{center}
\includegraphics[width=0.8\textwidth]{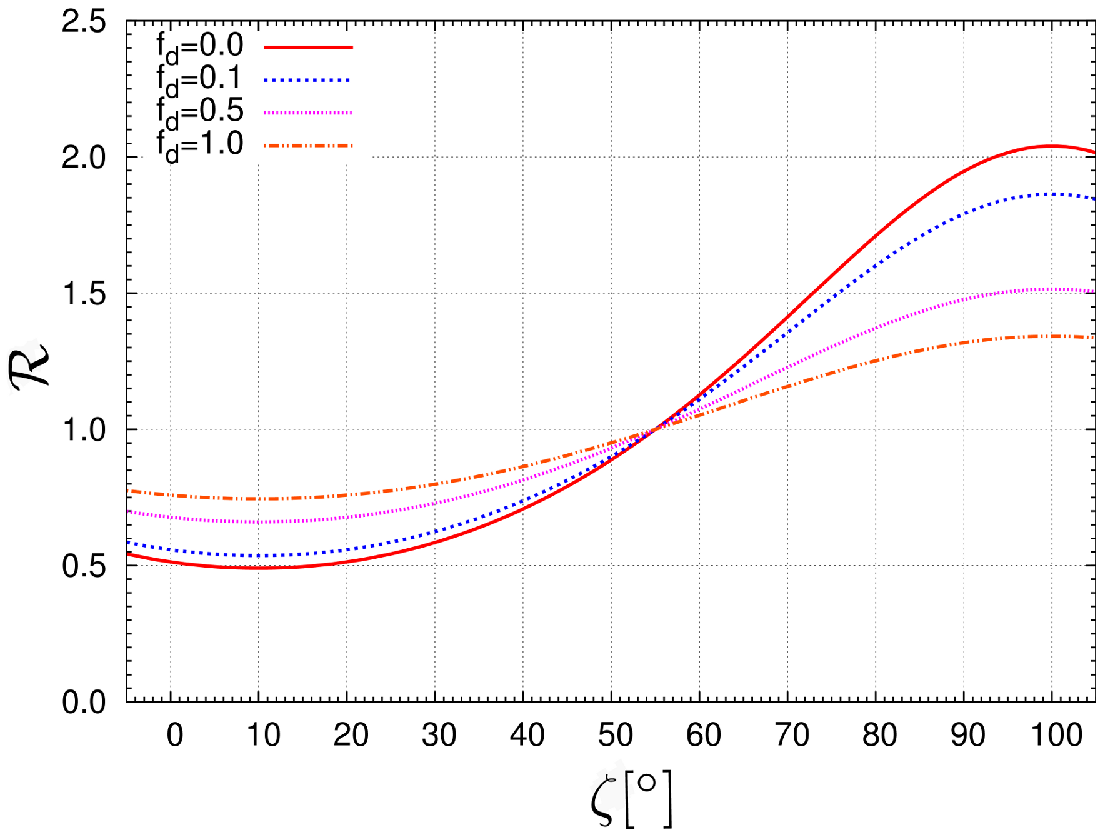}
\caption{Dependence on the power ratio $\mathcal{R}(\zeta)$ on the EBE diagnostic 
rotation angle. Different non-polarized background 
radiation fractions are considered}.
\label{fig:fig10}
\end{center}
\end{figure}

By rotating the diagnostic we should be able to determine if there is a dominant 
polarization in the incoming radiation and if this coincides with the expected one. 
For the ideal case of a perfect elliptically polarized wave coming from the 
B--X--O emission, the power dissipated along each detection direction when the diagnostic 
rotation angle is set to some arbitrary $\zeta$ is given by eqs. \ref{p1p2} of 
the Appendix. 
In order to consider 
the presence of an unknown amount of non-polarized radiation we define

\begin{equation}
\mathcal{R}(\zeta)\equiv\frac{P_1(\zeta)+P_d}{P_2(\zeta)+P_d}
\end{equation}
where $P_d\equiv f_dP_2(\zeta=10^\circ)$ is taken as a fraction of the maximum amplitude 
measured in channel 2. 
A minimum in $\mathcal{R}(\zeta)$ is expected for $\zeta=10^\circ$, when channel 2 is directed along the major axis of the polarization ellipse. 
The dependence of $\mathcal{R}$ on the rotation angle $\zeta$ of the horn antenna is represented in figure \ref{fig:fig10}
for different values of $f_{d}$.
As it is clear, 
the greater the amount of unpolarized radiation, the flatter the $\mathcal{R}(\zeta)$ 
profile becomes.
% SECTION BLOCK SECTION BLOCK SECTION BLOCK SECTION BLOCK SECTION BLOCK SECTION BLOCK
% SECTION BLOCK SECTION BLOCK SECTION BLOCK SECTION BLOCK SECTION BLOCK SECTION BLOCK
% SECTION BLOCK SECTION BLOCK SECTION BLOCK SECTION BLOCK SECTION BLOCK SECTION BLOCK
\section{Experimental results}
\label{exp_results}

\begin{figure}
\begin{center}
\includegraphics[width=0.8\textwidth]{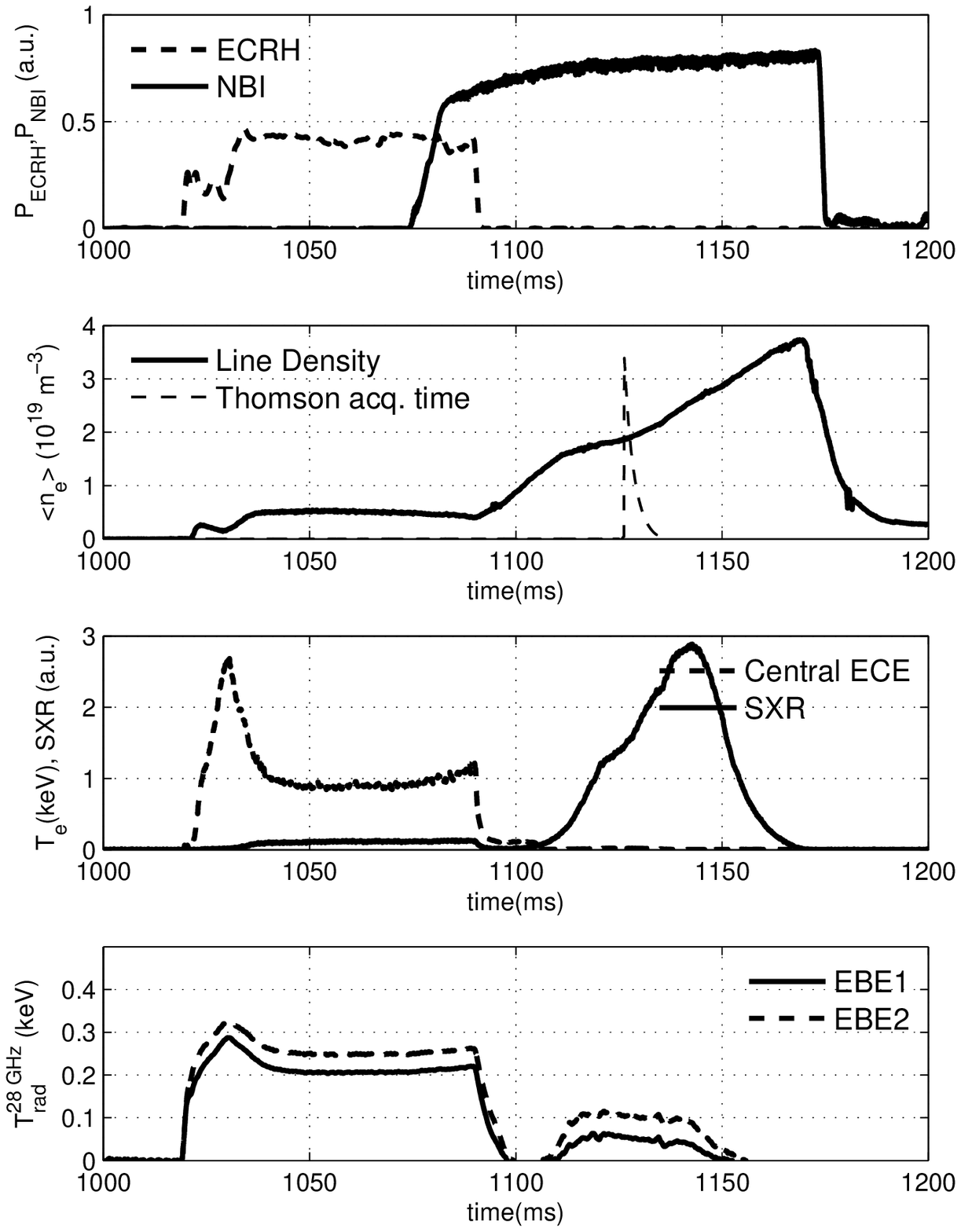}
\caption{Typical time evolution of some relevant quantities obtained in ECRH+NBI heated plasmas.}
\label{fig:fig11}
\end{center}
\end{figure}

\begin{figure}[t]
\begin{center}
\includegraphics[width=0.8\textwidth]{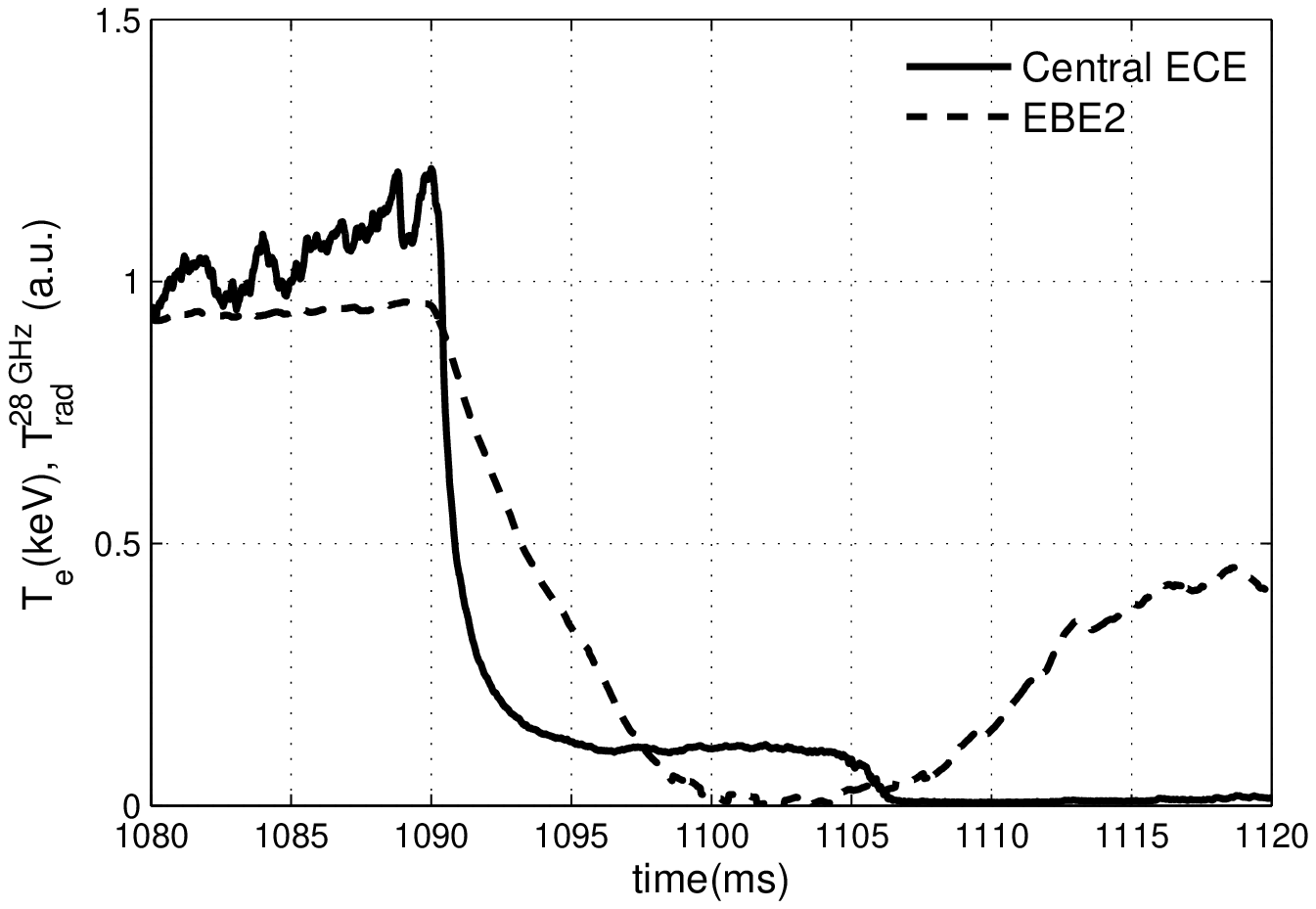}
\caption{Zoom view of the central ECE and EBE2 signals represented
in figure \ref{fig:fig11}.}
\label{fig:fig12}
\end{center}
\end{figure}

Figure \ref{fig:fig11} shows the time evolution of the main plasma parameters 
observed during a typical ECRH+NBI shot. 
Plasma heating is first performed by applying 500 kW of 
second harmonic off-axis ECRH power. Then, NBI heating ($P_{\text{\tiny{NBI}}}\approx$ 1 MW) 
is turned on and densities in the range $2-3.5\times 10^{19}$ m$^{-3}$ are achieved.
ECRH power is turned off before the second harmonic cut-off density 
for 53.2 GHz ($n_e\approx 1.7\times 10^{19}$ m$^{-3}$) is attained. 
This provokes the drop in the ECE signals, 
which still show a low plasma temperature until the NBI-induced electron density growth prevents the detection by ECE means.
The increase in the electron temperature and 
density during the NBI phase is accompanied with the subsequent
raise of radiation observed by the Soft X ray (SXR) detector.
EBE and ECE radiation after the ECRH turn-off 
are compared in figure \ref{fig:fig12}. 
The ECE signal decreases faster because this diagnostic is looking to the plasma along
a perpendicular 
line of sight whereas the EBE diagnostic is looking with a highly 
oblique view.
Thus, the 28 GHz emission comes
from suprathermal 
electrons with a lower collisionality than that of the thermal 
electrons responsible of the emission detected by the ECE diagnostic. 

\subsection{Experimental $\alpha_{\phi}$ and $\alpha_{\theta}$ scans}
\label{sec:scans}

\begin{figure}[t]
\begin{center}
\includegraphics[width=0.8\textwidth]{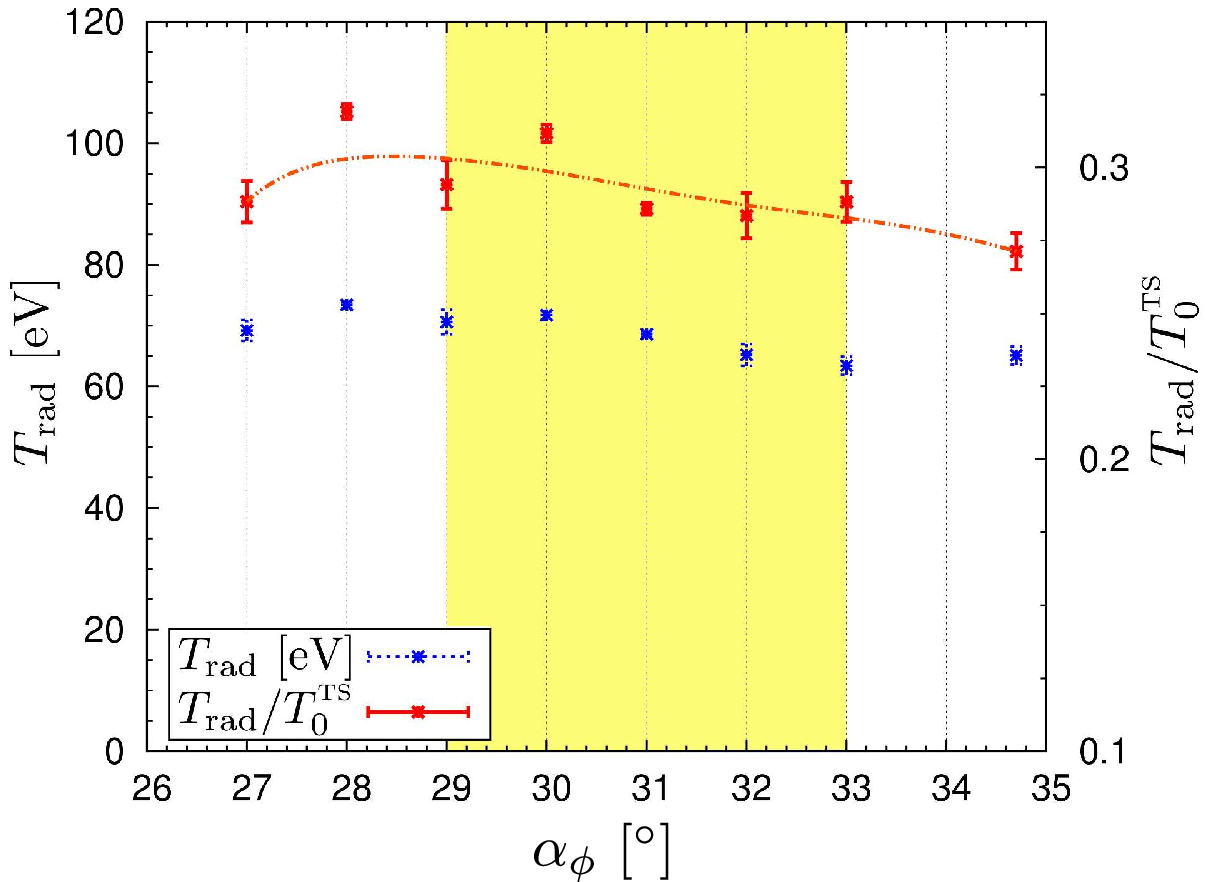}
\caption{Dependence of $T_{\text{rad}}$ 
on $\alpha_{\phi}$ 
for $\alpha_{\theta}=21.5^{\circ}$. The shadowed 
area represent the interval considered in the theoretical emission plots.}
\label{fig:fig13}
\end{center}
\end{figure}

\begin{figure}[t]
\begin{center}
\includegraphics[width=0.8\textwidth]{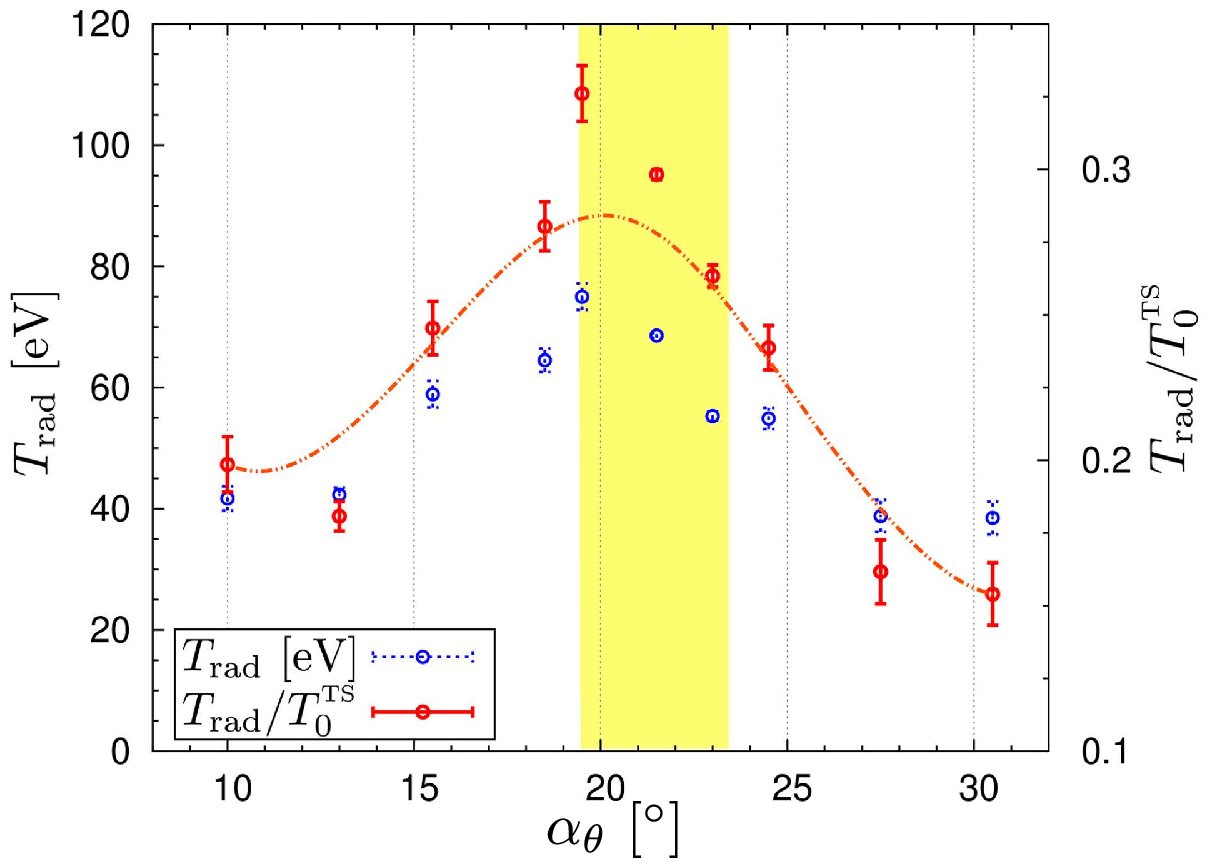}
\caption{Dependence of $T_{\text{rad}}$ on $\alpha_{\theta}$ 
for $\alpha_{\phi}=31^{\circ}$.}
\label{fig:fig14}
\end{center}
\end{figure}

Figures \ref{fig:fig13} and \ref{fig:fig14} show the measured radiative 
temperature at the TS acquisition time %EBE intensity 
for different positioning angles of the internal mirror. 
The first of them corresponds to a scan in $\alpha_{\phi}$
for $\alpha_{\theta}=21.5^{\circ}$, while the second
one represents, for $\alpha_{\phi}=31^{\circ}$, 
the scan in $\alpha_{\theta}$. 
In both figures $T_{\text{\tiny{rad}}}$ is referred to the left axis (blue dots). 
In order to remove its strong dependence on
the temperature profile,
which was not constant between shots,
these values are normalized to the central temperature value ($T_{0}^{\text{\tiny{TS}}}$) provided by
the TS diagnostic (red dots, right axis).

The scanned positions cover a wider area than the one considered in the theoretical 
calculations. The result presented in figure \ref{fig:fig13} is consistent with the theoretical 
predictions. The measured intensity show an almost flat profile 
weakly dependent on $\alpha_{\phi}$.
Regarding the scan shown in figure \ref{fig:fig14} a stronger variation of the
collected 28 GHz intensity on $\alpha_{\theta}$ is clearly manifested, 
as was also expected. 
The value for which a maximum emission value is obtained 
is $\alpha_{\theta}\approx19.5^{\circ}$. 
This can be interpreted as an acceptable result compared with the 
theoretical prediction ($\alpha_{\theta}\approx21.5^{\circ}$), considering that 
the range of movement of the mirror along the poloidal direction is $30^{\circ}$ approximately.
On the other hand this disagreement of approximately $2^{\circ}$ 
represents a noticeable deviation compared with the width of the EBE
window along the poloidal direction, which has $4^{\circ}$ (see 
figures \ref{fig:fig6}).
Furthermore, if all the radiation received came only from B--X--O-converted one,
 the intensity should drop drastically out of the $4^{\circ}$ of 
width of the viewing window. Since this is not what happens, 
it can be concluded that the amount of $28$ GHz disperse 
radiation collected by the mirror is far from being negligible. 
Such a contribution of disperse radiation is probably coming from EBE
emitted by the over dense plasma core all around the torus. This radiation leaves the plasma
via B--X--O or B--X conversion and reaches the antenna after reflecting on the 
walls, which also provoke randomly changes on its polarization.

\subsection{Experimental determination of polarization}
\label{rotationexp}

\begin{figure}[t]
  \begin{centering}
  \includegraphics[width=0.8\textwidth]{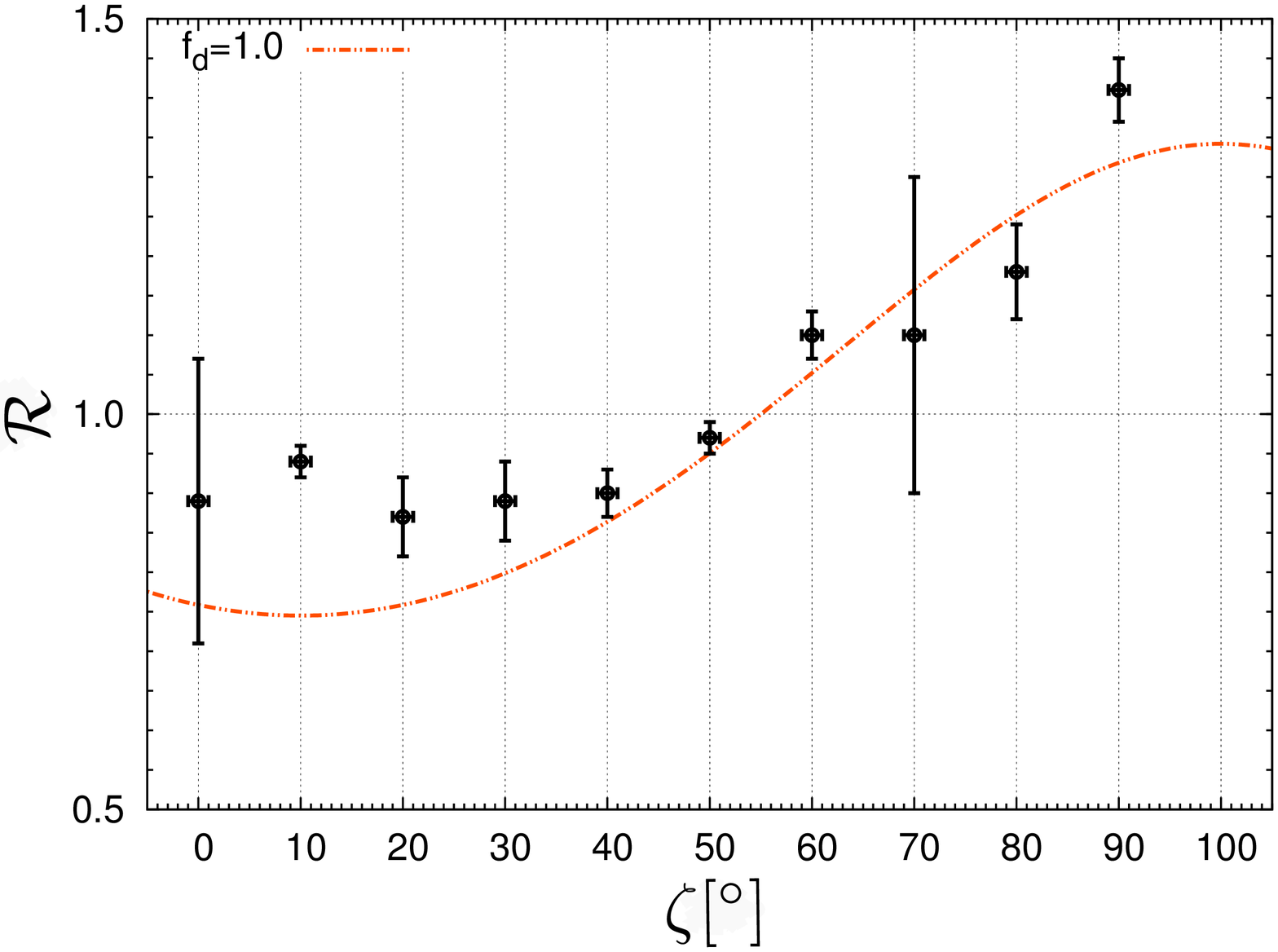}
  \caption{Experimental ratio $\mathcal{R}(\zeta)=I_{\text{\tiny{EBE1}}}/I_{\text{\tiny{EBE2}}}$ compared to 
the theoretical estimation, for $f_{d}=1$. A significant fraction of
disperse radiation is found.}
  \label{fig:fig15}
  \end{centering}
\end{figure}

A scan of the rotation angle of the diagnostic has been carried out 
in order to verify that the radiation reaching the mirror  contains a 
direct contribution of O-mode polarized radiation.
If this contribution exists, we
should observe a similar trend in the experimental and the theoretical ratio
$\mathcal{R}(\zeta)$ (see fig. \ref{fig:fig10}). 
The polarization
has been measured along the line of sight of the mirror for which the measured EBE intensity
is maximum along the poloidal angle (see fig. \ref{fig:fig14}). 
This line of sight corresponds to the values $\alpha_{\theta}=19.5^{\circ}$ and
$\alpha_{\phi}=31^{\circ}$.
The rotation angle $\zeta$ was varied from $-10^{\circ}$ to $90^\circ$. 
Figure \ref{fig:fig15} shows the experimental ratio 
$\mathcal{R}(\zeta)=I_{\text{\tiny{EBE1}}}(\zeta)/I_{\text{\tiny{EBE2}}}(\zeta)$ measured at 
the TS acquisition time and the calculated one when $f_{d}=1$, that is, the value for 
which the best agreement is observed.  
Taking also into account the result shown in fig. \ref{fig:fig14}, it can be concluded
that the background unpolarized field contributing
to the measured intensity is of the same order
of magnitude of the O-mode radiation reaching the detector.

% SECTION BLOCK SECTION BLOCK SECTION BLOCK SECTION BLOCK SECTION BLOCK SECTION BLOCK
% SECTION BLOCK SECTION BLOCK SECTION BLOCK SECTION BLOCK SECTION BLOCK SECTION BLOCK
% SECTION BLOCK SECTION BLOCK SECTION BLOCK SECTION BLOCK SECTION BLOCK SECTION BLOCK
\section{Summary and discussion}
\label{conclusions}

In this work a comprehensive study of the 28 GHz radiation 
rising from B--X--O conversion in over-dense TJ--II NBI-heated 
plasmas has been undertaken theoretically and experimentally.
The numerical results have been obtained with the TRUBA ray tracing code. 
As expected from an X--O conversion process, highly dependent on the wave propagation direction, a 
strong variation of the intensity has been found as the viewing angle along the poloidal direction is modified.
It has been observed that the experimental maximum of the emitted radiation
was shifted $1.5-2.0$ degrees with respect to
the numerical estimations, which is not negligible compared 
with the approximately 4 degrees
of width of the EBE window along the poloidal direction.
As a task for a forthcoming work, a fine scan on this
angle should be performed in order to resolve that maximum. Regarding the maximum value of $I_{\text{\tiny{EBE}}}$ 
along the toroidal direction, the observation is in agreement with the theoretical
prediction, since both theory and experiment 
shows an almost flat profile $I_{\text{\tiny{EBE}}}(\alpha_{\phi})$.
From these results, an optimum experimental direction for forthcoming EBW heating 
experiments is extracted.

\begin{figure}[t]
  \begin{center}
  \subfigure{\includegraphics[width=0.41\textwidth,angle=0]{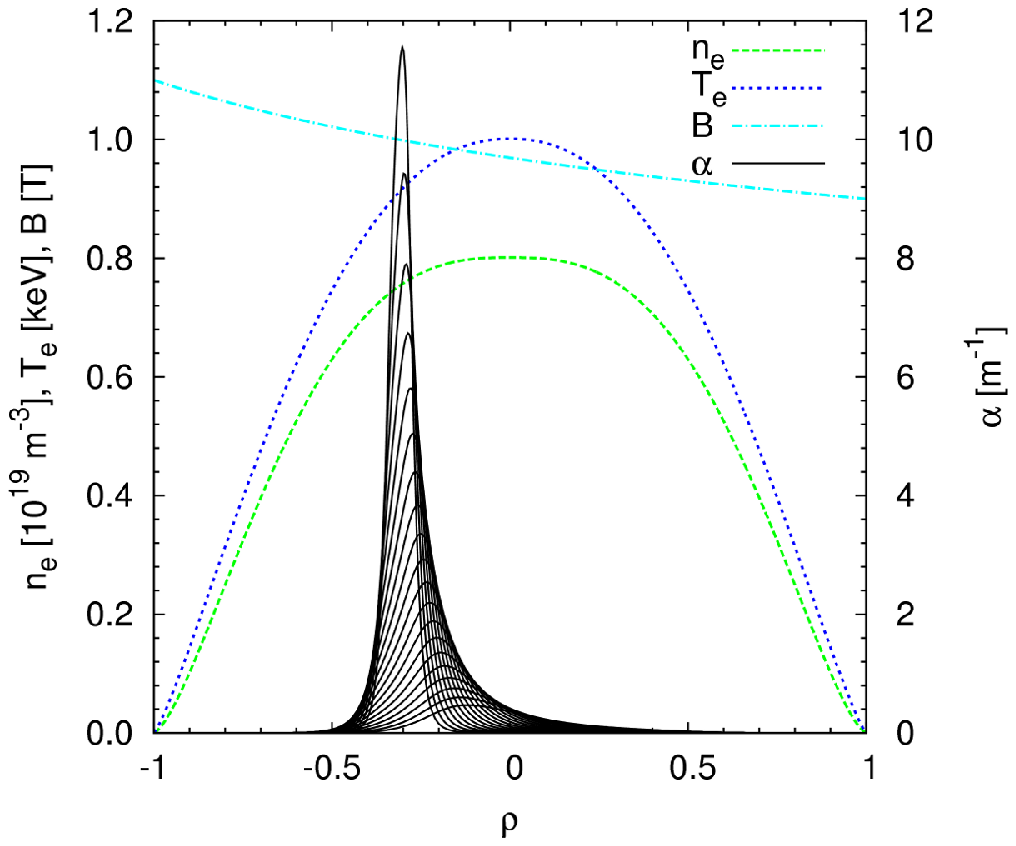}}~(a)
  \subfigure{\includegraphics[width=0.43\textwidth,angle=0]{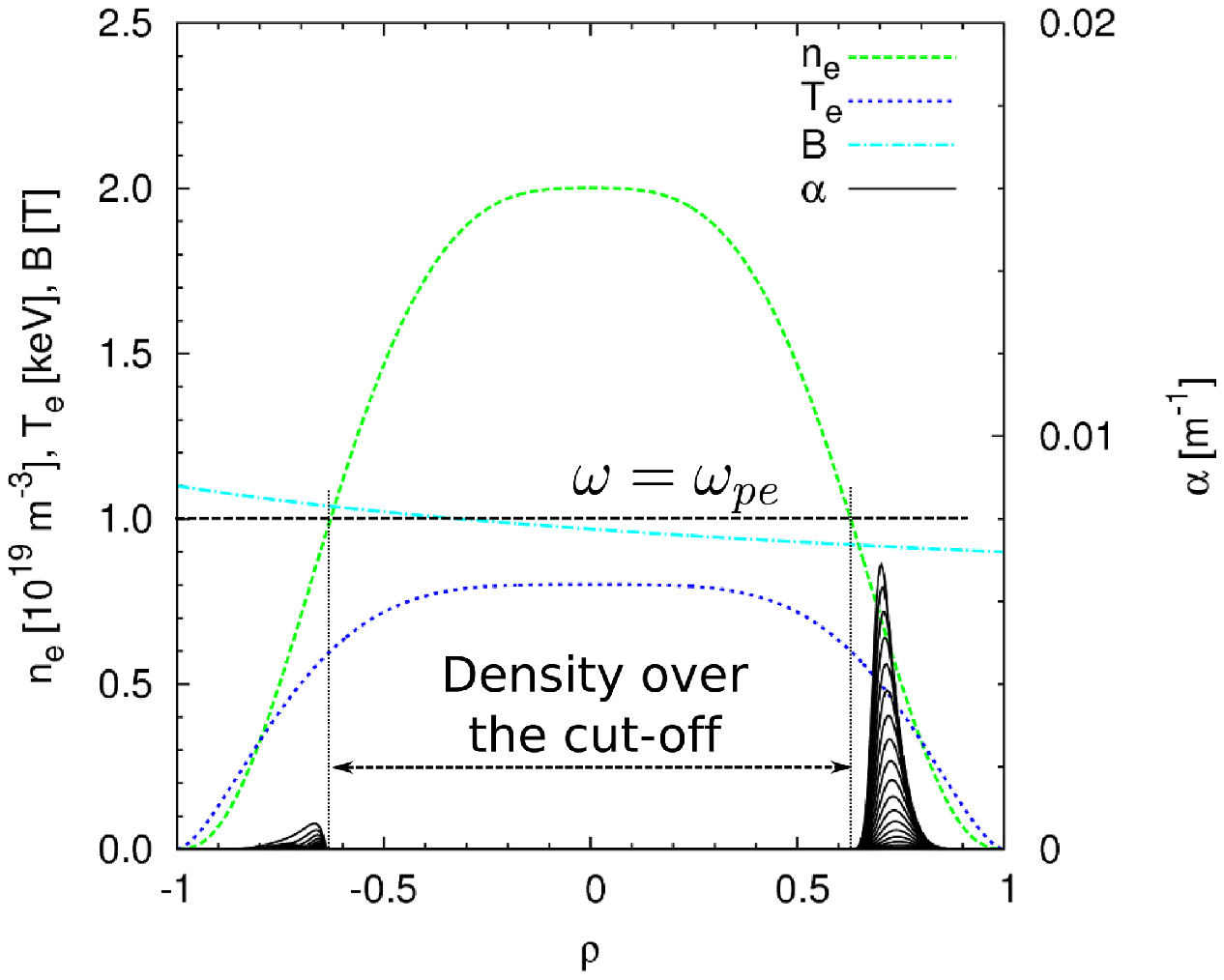}}~(b)
\caption{(a) Absorption coefficient (solid lines) for oblique O-mode propagation at $28$ GHz as a function of the effective radius, considering a cylindrical plasma with the magnetic field profile similar to the TJ-II one. The electron temperature (dashed line) and density (dotted line) profiles like those generated in ECRH-heated TJ-II plasmas are also represented. The density is under the O-mode cut-off in all the profile. (b) Oblique O--mode absorption coefficient for the over-dense case, i.e. profiles from NBI-heated TJ-II plasmas are considered. In both cases the different values of $N_{\|}$ considered are in the range between $N_{\|}\approx0$ and $N_{\|}\approx 0.65$. The more peaked curves of the absorption coefficient correspond to the lower value of $N_{\|}$.}\label{fig16}
\end{center}
\end{figure}

The polarization of the collected radiation has been measured in order
to verify that the received radiation comes mainly from a
B--X--O conversion process. The experiment has confirmed the existence of 
the expected O-mode component, importantly screened by disperse radiation.

Further discussion can be made on the possibility that this O-mode 
could be emitted from the under-dense plasma edge. 
This point is clarified in figs. \ref{fig16}(a) and \ref{fig16}(b), 
which show the O-mode absorption coefficient
for different values of $N_{\|}$ in a cylindrical
plasma with typical TJ--II magnetic field, electron density and temperature profiles.
This has been estimated by using well-known
expressions of the weakly relativistic absorption coefficient of electromagnetic
modes \cite{Bornatici_nf_23_1983}.
On the one hand, figure \ref{fig16}(a) represents the case where the density is below the 
O-mode cut-off, i.e. between 1025 and 1100 ms in figure \ref{fig:fig11}. 
On the other hand, figure \ref{fig16}(b) shows the over-dense
case in the interval between 1100 and 1150 ms.
The absorption coefficient in the over-dense scenario
is three orders of magnitude lower than in the under-dense case. Since in this situation the absorption
of the O-mode is negligible in the edge, so is the emission coefficient. 
This allows to conclude that the O mode polarized radiation measured when in over-dense 
conditions can only be attributed to a B--X--O process.

A considerable amount of 
disperse radiation has been found. Its presence can be understood by considering that
the emission of EBWs takes place all around the over-dense plasma column 
and reaches the antenna after many reflections in the vacuum chamber, which in
turn produces a random polarization. Further studies on 
the characterization of the disperse radiation using ray tracing are planned.
This task results extremely important due to the role of this 
contribution to the deterioration of the EBE signal. 
As a matter of fact, the disperse radiation together with
the presence of the X--O conversion layer in between 
the emitting source and the radiometer horns, represents
a handicap in the calibration of the EBE signal by 
traditional means if the electron temperature diagnosis in over-dense regime is aimed. Note that 
X--O conversion layer act as a kind of $\mathbf{k}$-spectrum filter that can not be 
taken into account in the calibration procedure. It could only be included a posteriori 
knowing the exact O--X conversion efficiency of the beam and assuming a 
symmetric conversion efficiency, which is not always true. Moreover, as it was briefly mentioned in 
section \ref{num_results}, the calculation of the beam O-X conversion efficiency by ray tracing is not accurate, making the final determination 
of an electron temperature a very hard task.

Finally, the improvement in the spatial resolution by considering a more collimated 
viewing pattern as in standard ECE measurements could help to get rid of the disperse radiation. However, this will be accompanied by a X--O
conversion efficiency drop due to the widening of the $\mathbf{k}$-spectrum of the field 
associated to a highly collimated beam and, consequently, the EBE source 
for the temperature measurement would be strongly screened by the disperse radiation.

% SECTION BLOCK SECTION BLOCK SECTION BLOCK SECTION BLOCK SECTION BLOCK SECTION BLOCK
% SECTION BLOCK SECTION BLOCK SECTION BLOCK SECTION BLOCK SECTION BLOCK SECTION BLOCK
% SECTION BLOCK SECTION BLOCK SECTION BLOCK SECTION BLOCK SECTION BLOCK SECTION BLOCK
\section*{Acknowledgments } 
This work has been partially funded by the Spanish Ministerio de Ciencia e Innovaci\'on, Spain, 
under Project ENE2008-06082/FTN

\section*{Appendix: Expected power for an arbitrary rotation angle}
\label{powratio}
\begin{figure}[t]
  \begin{centering}
  \includegraphics[width=0.5\textwidth]{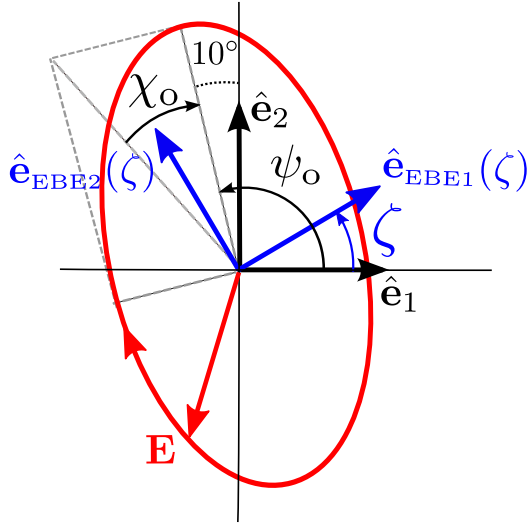}
  \caption{Theoretical O-mode polarization ellipse. In the figure, the azimuth is defined in the non rotated frame ($\zeta=0$). The polarizations vectors of the detector when it is rotated an angle $\zeta$, $\hat{\mathbf{e}}_{\text{\tiny{EBE1}}}(\zeta)$ and $\hat{\mathbf{e}}_{\text{\tiny{EBE2}}}(\zeta)$, are also represented.}
  \label{fig:fig17}
  \end{centering}
\end{figure}

For the derivation 
of an explicit form of the powers $P_{1}(\zeta)$ and $P_{2}(\zeta)$, 
the ideal theoretical polarization of the B--X--O radiation is first defined
in the main reference system given by the polarization vectors
$\hat{\mathbf{e}}_{1}$ and $\hat{\mathbf{e}}_{2}$ (see figure \ref{fig:fig17}). 
As already anticipated, the azimuth of the O mode polarization
ellipse in this reference system can be obtained from geometrical calculations
by finding the projection of the $\{\mathbf{k},\mathbf{B}\}$ plane, defined 
at the plasma boundary, onto the detection plane of the horn, after 
the two successive reflections both in the elliptical and the flat mirror.
Furthermore, in order to couple an obliquely propagating ordinary mode at
the plasma edge the wave must be elliptically polarized with an ellipticity
angle $\gamma$ given by
\begin{equation}
\tan{\gamma}=\frac{Y(\sin^{2}\theta-\varrho)}{2\cos\theta},
\end{equation}
where $\varrho^2\equiv \sin^{4}\theta+(4/Y^{2})\cos^{2}\theta$, $Y=\omega_{ce}/\omega$ 
and $\theta$ is the propagation angle at the plasma edge.
For the standard TJ--II magnetic
configuration and taking into account the fact that the detector is looking
to the emitted wave (the rotation sense of the electric field is reversed), we
finally obtain the angles that define the polarization ellipse in the 
main reference system, i.e.
\begin{equation}
\label{elipseparam}
\psi_{\text{\tiny{O}}}=100^{\circ},\ \ \
\chi_{\text{\tiny{O}}}\equiv-\gamma=-35^{\circ}.
\end{equation}
Taking $|E_{1}|^2+|E_{2}|^2\equiv 1$ the relation between these angles and 
the wave electric field is given by \cite{Chandrasekhar_radiative}
\begin{equation}
\label{stokes}
\begin{split}
&s_{1}=E_{1}^2-E_{2}^2=\cos 2\chi_{\text{\tiny{O}}}\cos 2\psi_{\text{\tiny{O}}},\\
&s_{2}=2|E_{1}||E_{2}|\cos\delta=\cos 2\chi_{\text{\tiny{O}}}\sin 2\psi_{\text{\tiny{O}}},\\
&s_{3}=2|E_{1}||E_{2}|\sin\delta=\sin 2\chi_{\text{\tiny{O}}},
\end{split}
\end{equation}
where $|E_{1}|$, $|E_{2}|$ and $\delta\equiv\delta_{2}-\delta_{1}$ determine 
the complex electric field amplitude through $\mathbf{E}=E_{1}\hat{\mathbf{e}}_1+E_{2}\hat{\mathbf{e}}_{2}=
|E_{1}|\exp(\text{i}\delta_{1})\hat{\mathbf{e}}_{1}+
|E_{2}|\exp(\text{i}\delta_{2})\hat{\mathbf{e}}_{2}$.
Consider now the field components in an arbitrary $\zeta-$rotated system defined 
by the two orthogonal directions of the EBE diagnostic: $\mathbf{E}=E'_{1}\hat{\mathbf{e}}_{\text{\tiny{EBE1}}}+E'_{2}\hat{\mathbf{e}}_{\text{\tiny{EBE2}}}$.Note that when $\zeta\approx10^{\circ}$, the vector $\hat{\mathbf{e}}_{\text{\tiny{EBE2}}}$ is directed 
along the major axis of the polarization ellipse. 
Therefore, the power detected in both channels, for an arbitrary angle $\zeta$, is given by 

\begin{equation}
\begin{split}
&P_{1}(\zeta)=E'_{1}{E'_{1}}^{*}=|E_{1}|^{2}\cos^{2}\zeta+
|E_{2}|^{2}\sin^{2}\zeta+2\Re(E_{1}E_{2})\cos\zeta\sin\zeta\\
&P_{2}(\zeta)=E'_{2}{E'_{2}}^{*}=|E_{2}|^{2}\cos^{2}\zeta+
|E_{1}|^{2}\sin^{2}\zeta-2\Re(E_{1}E_{2})\cos\zeta\sin\zeta.
\end{split}
\end{equation}
Recovering the relations (\ref{stokes}) and taking $\delta_1=0$, $P_{1}$ 
and $P_{2}$ can be written as

\begin{equation}
\label{p1p2}
\begin{split}
&P_{1}(\zeta)=\left(\frac{1+s_{1}}{2}\right)\cos^{2}\zeta+\left(\frac{1-s_{1}}{2}\right)
\sin^{2}\zeta+s_{2}\cos\zeta\sin\zeta\\
&P_{2}(\zeta)=\left(\frac{1-s_{1}}{2}\right)\cos^{2}\zeta+\left(\frac{1+s_{1}}{2}\right)
\sin^{2}\zeta-s_{2}\cos\zeta\sin\zeta.
\end{split}
\end{equation}
For $\zeta=10^{\circ}$, $P_{2}$ is maximum.

% SECTION BLOCK SECTION BLOCK SECTION BLOCK SECTION BLOCK SECTION BLOCK SECTION BLOCK
% SECTION BLOCK SECTION BLOCK SECTION BLOCK SECTION BLOCK SECTION BLOCK SECTION BLOCK
% SECTION BLOCK SECTION BLOCK SECTION BLOCK SECTION BLOCK SECTION BLOCK SECTION BLOCK

\end{document}